\documentclass[11pt,preprint2]{aastex}

\usepackage{pdfpages}
\usepackage{natbib}
\usepackage{graphicx}
\begin{document}

% The following seven commands are intended for editorial usage and should be ignored by
% the author(s).
%\Pagespan{789}{}% Document's page range. 
% If second parameter is left empty, the last page is computed automatically.
%\Yearpublication{2013}%
%\Yearsubmission{2013}%
%\Month{11}%   
%\Volume{999}%  
%\Issue{88}% 
% \DOI{This.is/not.aDOI}% 

\title{Intrabinary Shock Emission from Black Widows and Redbacks}
\author{Roberts, Mallory S.E.}
\affil{Eureka Scientific, Inc., Oakland, CA, USA and New York University Abu Dhabi, UAE }
%\altaffiltext{1}{New York University Abu Dhabi, P.O. Box 129188, Abu Dhabi, United Arab Emirates}
\author{McLaughlin, Maura A. and Gentile, Peter}
\affil{West Virginia University, Morgantown, WV, USA}
\author{Aliu, Ester}
\affil{Barnard College, Columbia University, New York, NY, USA}
\author{Hessels, Jason W.T.}
\affil {ASTRON, Dwingeloo, the Netherlands}
\author{Ransom, Scott M.}
\affil{National Radio Astronomy Observatory (NRAO), Charlottesville, VA, USA}
\and
\author{Ray, Paul S.}
\affil{Naval Research Laboratory, Washington, DC, USA}
\shorttitle{Black Widows and Redbacks}
\shortauthors{M.S.E. Roberts et al. }
%\institute{
%Eureka Scientific
%Oakland, CA. USA
%\and 
%New York University Abu Dhabi, U.A.E.
%\and 
%West Virginia University, USA
%\and
%Barnard University, USA
%\and
%ASTRON, Netherlands
%\and
%NRAO, USA
%\and
%Naval Research Lab, USA
%}

%\received{16 Sep 2013}
%\accepted{13 Oct 2013}
%\publonline{later}
%%\maketitle

%\keywords{pulsars: general, gamma rays: observations, radio continuum: stars, X-rays: binaries, shock waves}

\begin{abstract}
Eclipsing millisecond pulsars in close ($P_b < 1$~day) binary systems provide a different view of pulsar winds and shocks than do isolated pulsars. Since 2009, the numbers of these systems known in the Galactic field has increased enormously. We have been systematically studying many of these newly discovered systems at multiple wavelengths. Typically, the companion is nearly Roche-lobe filling and heated by the pulsar which drives mass loss from the companion. The pulsar wind shocks with this material just above the surface of the companion. We discuss various observational properties of this shock, including radio eclipses, orbitally modulated X-ray emission, and the potential for $\gamma$-ray emission. Redbacks, whose companions are likely non-degenerate and significantly more massive, generally have more luminous shocks than black widows which have very low mass companions. This is expected since the more massive redback companions intercept a greater fraction of the pulsar wind. We also compare these systems to accreting millisecond pulsars, which may be progenitors of black widows and in some cases can pass back and forth between redback and accretion phases. 
\end{abstract}

\section{Introduction}

In 1986, the first binary pulsar to exhibit radio eclipses, PSR B1957+20, was discovered at the Arecibo observatory (Fruchter et al. 1988a). This 1.61~ms pulsar is in a 9.2~hr orbit around a very low mass ($M_c \sim 0.02 M_{\odot}$) companion. Regular radio eclipses were observed for $\sim 10\%$ of the orbit when the companion was at inferior conjunction. During the ingress and egress of the eclipse, delays in the pulse arrival times were observed which, along with the eclipse duration, suggested that the eclipses were due to excess ionized gas within the system (although the exact physical mechanism of the eclipses was and still is unclear eg. Thompson 1995). Optical studies showed large orbital variations in the light from the companion, suggesting the emission was primarily due to illumination by the pulsar (Fruchter et al. 1988b). H$\alpha$ (Kulkarni et al. 1988) and X-ray (Stappers et al. 2003) imaging showed unambiguous evidence for a pulsar wind nebula around the pulsar, and observations of the non-thermal X-ray point source showed evidence for orbital modulation as well (eg. Huang and Becker 2007). Long term timing has shown that there are orbital period variations which are much larger in magnitude than would be expected from gravitational wave emission, with an orbital period derivative that can even change sign over time (Arzoumanian et al. 1994). These have  no easy interpretation, but mass loss from the system and an induced quadrapole moment of the companion could contribute to the measured changes in orbit (Applegate \& Shaham 1994).

The inference drawn from these various observations is that PSR B1957+20 is ablating material off of its companion which shocks with the pulsar wind. Such systems had  been proposed as a stage in the formation of isolated millisecond pulsars by Ruderman et al. (1989). Since the pulsar seemed to be in the process of destroying its companion, it was dubbed the ``black widow" pulsar. Early models of the shock emission expected from the original Black Widow pulsar were developed by Arons and Tavani (1993) and Raubenheimer et al (1995). These predicted electrons could be accelerated to $\sim 3$ TeV and that orbital modulation could arise from obscuration by the companion of part of the shock, intrinsic beaming of the emission by the magnetic field in the shock, or by doppler boosting. The high energy luminosity could depend on the distance to the shock, the magnetic field of the pulsar, the ion fraction, the optical emission of the companion (in the case of inverse Compton emission), and the magnetization of the wind. However, the overall scaling should be primarily due to spin-down power of the pulsar $\dot E$, the fraction of the wind involved in the shock $f$, and the distance to the system $d$. 

\begin{equation}
L_S\propto f \dot E/d^2 
\end{equation}

No unambiguous detection of $\gamma$-ray emission arising from the shock in PSR B1957+20 has yet been made (although there has been an intriguing suggestion of GeV emission in $Fermi$ data, see Wu et al. 2012). However, the ability of a pulsar wind to produce emission from radio frequencies up to TeV energies in an intrabinary shock has been demonstrated by observations of the PSR B1259$-$63 system (\cite{jmm+99,aaa+11,hess13}) which consists of a young pulsar in an highly eccentric 3.4 year orbit around a main sequence Be star. Near periastron the radio pulses of PSR B1259$-$63 are obscured as it enters the region of the dense equatorial outflow of the Be star, and unpulsed radio, X-ray, GeV and TeV emission is observed. While the $\dot E$ of PSR B1259$-$63 is somewhat higher and a larger fraction of the pulsar wind is shocked by the stellar wind than in PSR B1957+20, the total shock luminosity should not be much more than an order of magnitude greater. However, near periastron, PSR B1259$-$63 is a few hundred times brighter in X-rays than PSR B1957+20, despite being at similar distances.  

The big difference between the Black Widow shock and those in PSR B1259-63 and the termination shocks of young pulsar wind nebulae is the scale of the systems. In terms of light cylinder radii $R_{lc}$, the companion of the Black Widow pulsar is on the order of $10^4 R_{lc}$ rather than the more typical $10^8 R_{lc}$ to the inner torii of young pulsar wind nebulae. This forces the shock to be in the region where, if the pulsar wind is magnetically dominated at the light cylinder, the magnetization fraction $\sigma$ should still be relatively high. This may allow direct observational tests of proposed solutions to the ``sigma problem" of pulsar wind shocks where the inferred magnetization at the termination shock from models of the X-ray emission is very low despite the expectation that at the light cylinder it should be very high (\cite{a12}). 
Other potential advantages of studying the shocks in black widow type systems are that pulsar timing and optical observations allow us to determine the geometry of the system very well, and short orbits ($< 1$~d) allow for many repeated observations of the shock as viewed from different orientations. 

For many years it seemed that eclipsing systems in the Galactic field might be quite rare. In the 20 years following the discovery of PSR B1957+20, only 2 other systems were discovered, both of which had spin-down fluxes $\dot E/d^2$ much smaller than that of the original Black Widow (see Roberts 2011 and references therein). Observations of globular clusters, however, 
revealed that a significant fraction of their millisecond pulsars either eclipsed and/or had very low companion masses. Freire (2005) listed 18 such systems. 11 have very low mass companions ($M_c \la 0.04 M_{\odot}$)like PSR B1957+20, but 7 have masses more typical of helium white dwarfs. Optical observations of these latter systems tended to indicate the companions were, in fact, non-degenerate. Were these then systems where the pulsar was spun up by one companion, then exchanged the original companion for a non-degenerate one in the very crowded core of the globular cluster? If this were the case, then it would be natural that most such systems would be found in globular clusters. 

\section{PSR J1023+0038 the Canonical Galactic Field Redback}

During a drift scan survey for radio pulsars  with the Green Bank Telescope, a 1.69~ms pulsar was discovered in a 4.8~hr orbit around a $\sim 0.2 M_{\odot}$ companion coincident with the radio source FIRST J102347.6+003841 (Archibald et al. 2009). This source had been noted previously as a possible cataclysmic variable (\cite{bwb+02}) whose optical spectrum had undergone a drastic change $\sim 2001-2002$ (\cite{wat+09}) from a source with strong, broad, double-peaked emission lines implying an accretion disk to one more consistent with that of a G-type star with no emission lines but strong orbital modulation (\cite{ta05}). Modeling of the photometric light curve along with radial velocity measurements implied a hidden companion with a mass more in line with a neutron star that a white dwarf. The radio pulsations not only confirmed the neutron star nature of the primary, but also showed regular, frequency dependent orbital eclipses. This was reminiscent of both the Black Widow pulsar and the systems in globular clusters with non-degenerate companions, the first of which was discovered with the Parkes telescope. However this source is in the Galactic field and there was evidence for a recent accretion phase, suggesting that these systems with non-degenerate companions represented a phase in the life of system still undergoing the recycling process where the accretion is only temporarily interrupted and goes into a temporary phase where the freshly revealed radio pulsar ablates its companion. As more such systems were found in the Galactic field, they were given the moniker ``redback", after the Australian cousin to the North American black widow spider. 

Further optical studies and VLBI studies of PSR J1023+0038 combined with radio timing studies allowed for an accurate estimate of the intrinsic spin down, the distance, the inclination angle, and the companion's Roche lobe filling factor (\cite{dab+12}). They also showed the neutron star to be massive, likely to be $\ga 2 M_{\odot}$, implying a moment of inertia significantly larger than the canonical $10^{45} {\rm gm}\,{\rm cm}^2$ derived from an assumed mass of $1.4 M_{\odot}$ and radius of 10~km. PSR J1023+0038 so far is the only Galactic field redback with a VLBI determined parallax and proper motion as well as optical observations implying a massive neutron star with a Roche lobe filling companion. Since dispersion measure distances determined from the NE2001 free electron model (\cite{cl02}) 
often underestimate the true distance by as much as a factor of 2 for mid Galactic latitude pulsars (as was the case for PSR J1023+0038), and relative motion of the source to us can strongly affect the observed spin-down rate 
 through the Shklovskii effect (\cite{s70}) and the interaction region depends on the angle distended by the companion and hence the mass ratio and Roche lobe filling factor, any interpretation of shock emission from these systems should take into account these areas of uncertainty and ideally all would be observationally constrained. 

In the case of PSR J1023+0038, fairly extensive X-ray observations have been made showing clear evidence of emission from the shock (\cite{akb+10}). \cite{bah+11} have modelled the X-ray lightcurve and find that the evidence points towards an emission region very near the companion and that the orbital variations primarily come from partial obscuration of the shock region by the companion. Given the level of X-ray emission and assuming it is primarily synchrotron, they infer a fairly strong magnetic field of $\sim 40$~G at the shock. If this is the magnetic field carried by the wind to the shock, it would indicate a very high magnetization fraction in the wind, and that pulsar winds out to at least $\sim 10^4$ light cylinder radii are still magnetically dominated (i.e. have a high $\sigma$). An upper limit on the TeV flux from VERITAS supports a scenario with a moderately strong magnetic field at the shock (Aliu et al. in preparation). However, it is not clear that the inferred magnetic field is necessarily  due to the pulsar wind and not the magnetosphere of the companion. For example, if the dipole moment of the companion is similar to that of the Sun, then it's surface field would be on $\sim 25$~G, which could significantly influence the emission of electrons accelerated in a shock near the star's surface. PSR J1023+0038 has also been weakly detected by $Fermi$, but it is not yet clear if this emission is pulsed (\cite{thh+10}).  

\section{The New Population of Black Widows and Redbacks Associated with Fermi Sources}

Since the discovery of PSR J1023+0038, both large scale surveys and targeted searches of unassociated $Fermi$ $\gamma$-ray sources have resulted in a swarm of new black widow and redback discoveries (\cite{rob13}). Currently, there are $\sim 17$ black widows and $\sim 7$ redbacks known in the Galactic field.
Typically, they have spin-down powers (uncorrected for Shklovskii effect) $\dot E \sim 10^{34-35} {\rm erg}\,{\rm s}^{-1} $, distances between $\sim 0.5-3.5$~kpc (based on dispersion measures),
and orbital periods much shorter than a day. Those with long term timing solutions, especially redbacks, show orbital period variations similar to what is seen in the original black widow. The vast majority of these new systems are associated with $Fermi$ sources. However, in almost all cases once a good timing solution is obtained, pulsed GeV emission can be detected while there is little evidence for any $\gamma$-ray emission coming from the shock. The pulse period distribution is on average shorter than the general radio MSP population, and similar to what is seen in accreting X-ray millisecond pulsars (AXMSPs) (\cite{pw12}). Comparison of orbital period-companion mass relationship of redbacks and black widows to AXMSPs show that the latter appear to bridge the two populations. Accreting sources with similar companion masses to redbacks also have similar orbital periods, while those with companion masses similar to black widows tend to have shorter orbital periods. This suggests that accretion may turn on and off when the companion has not yet fully shed its non-degenerate envelope, and so can either be a redback or an AXMSP, but for very light companions, once the pulsar turns on the accretion phase has ended. The ability to pass back and forth between a redback phase and an AXMSP 
phase has been spectacularly demonstrated by the globular cluster redback M28I.
This system was observed to briefly go into an accretion phase and become an AXMSP in Spring 2013 (\cite{pfb+13}) and then turned back into a radio pulsar within a few months (\cite{phb+13}). While interruptions in the accretion process are expected on the evolutionary timescale of the companion, the mechanism for such rapid switching back and forth is currently not understood. 

\begin{figure}[h!]
\hspace*{-8mm}
\includegraphics[width=65mm,angle=90]{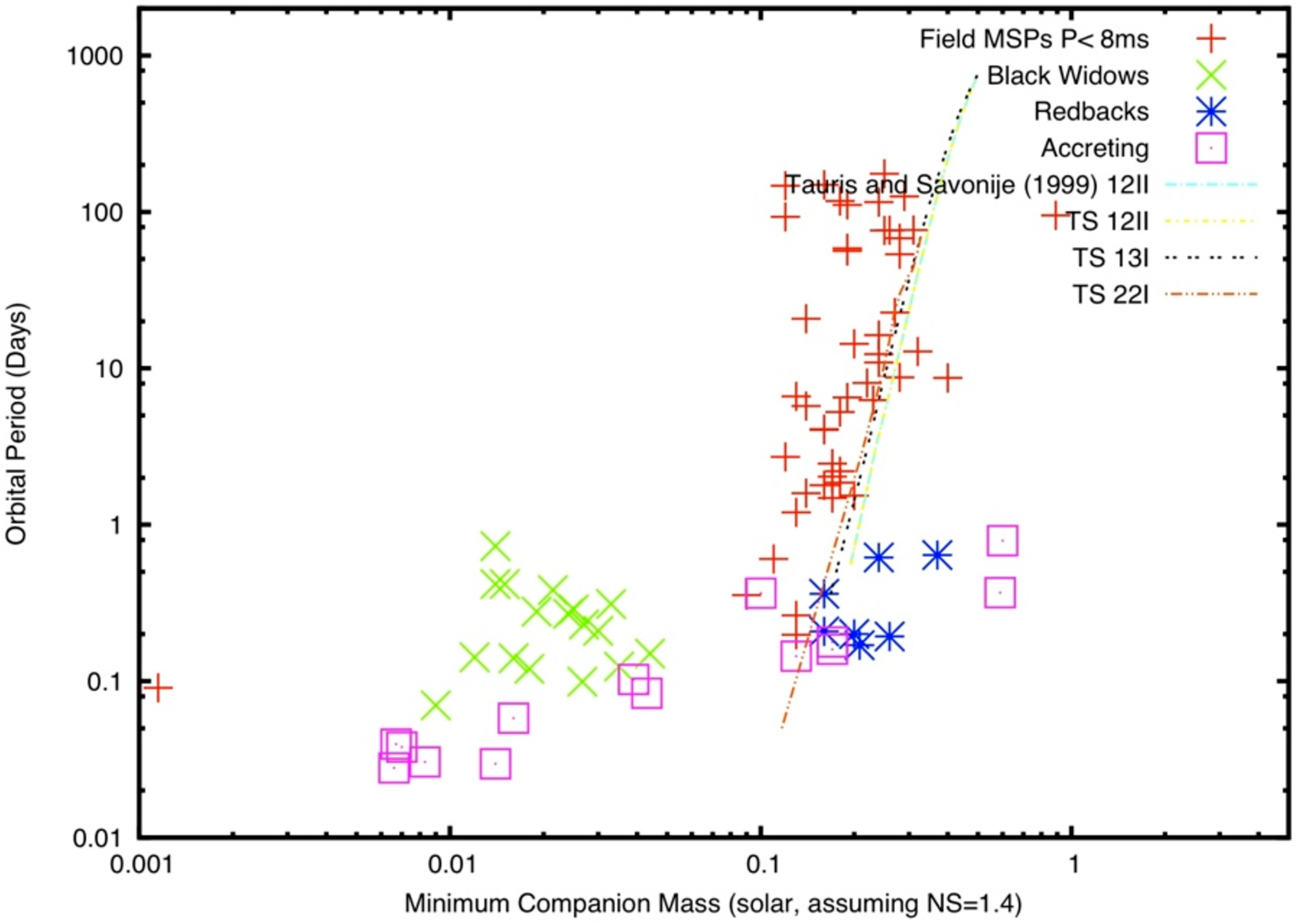}
\caption{Minimum companion mass vs. orbital period of fast ($P < 8$ ms) binary MSPs in
the Galactic field, showing the positions of Redbacks (blue stars), Black Widows (green crosses) and AXMSPs (pink boxes) (\cite{rob13,pw12}). 
The lines are from various binary evolution models considered by Tauris \& Savonije
(1999) which result in a Helium white dwarf companion. Since we plot minimum companion
masses, we expect systems which evolved according to standard evolutionary scenarios be just
to the left of these lines. 
Non-spider MSPs (plus signs) are from the ATNF pulsar
catalog http://www.atnf.csiro.au/research/pulsar/psrcat}
\label{label1}
\end{figure}

A number of the new systems have been studied in the optical. Most seem to be nearly Roche lobe filling, but at least one black widow (PSR J0023+0923) and possibly a second (PSR J2256$-$1024) significantly underfill their Roche lobes (\cite{bvr+13}). 
There is also one system (PSR J1816+4510) which shows eclipses and has a companion mass from timing $> 0.16 M_{\odot}$ like a redback (\cite{ksr+12}), but optical studies show that its companion is a very hot, metal rich, low-gravity star which significantly underfills its Roche lobe and is spectrally similar to a proto-white dwarf. It therefore may just be ending its redback phase (\cite{kbv+13}).

Several of the new systems have also had X-ray observations covering at least one orbit (\cite{grm+13}), although none have been as extensively studied in X-rays as PSR B1957+20 and PSR J1023+0038. Their X-ray spectra mostly seem to be a mixture of thermal and non-thermal emission. Some show strong evidence for orbital variation, while others show hints of orbital variation. An exception is PSR J0023+0923, which is in keeping with the companion underfilling its Roche lobe and hence intersect with a smaller fraction of the pulsar wind. In Figure~2, we plot the measured X-ray flux as a function of the quantity $(\dot E/d^2)(R_L/2a)^2$ where $R_L$ is the average Roche lobe radius of the companion and $a$ is the semi-major axis of the orbit. Hence this quantity should be proportional to the total potential flux from any intrabinary shock, assuming the companion fills its Roche lobe. Note we assume the standard formula for $\dot E$ and do not try to take into account differences in moment of inertia and in most cases there are no proper motion constraints and hence no correction for the Shklovskii effect. The distances, other than for PSR J1023+0038, are from the pulsar dispersion measure. For cases where there is evidence for underfilling of the Roche lobe and/or a different distance, we have indicated the effect on the inferred shock flux with arrows of varying length. The measured values do seem to roughly correlate  to the expected total flux from the intrabinary shock. Such a relationship will hopefully become clearer once more sources have more reliable estimates of their intrinsic spin-downs, true distances and Roche lobe filling factors, and also have good enough statistics in their X-ray spectra to clearly separate non-thermal flux emitted by the shocked wind from thermal emission emitted by the neutron star surface.  None of the newer sources has yet to show evidence of extended nebular emission as seen around PSR B1957+20.

\begin{figure}[h!]
\hspace*{-8mm}
\includegraphics[width=65mm,angle=90]{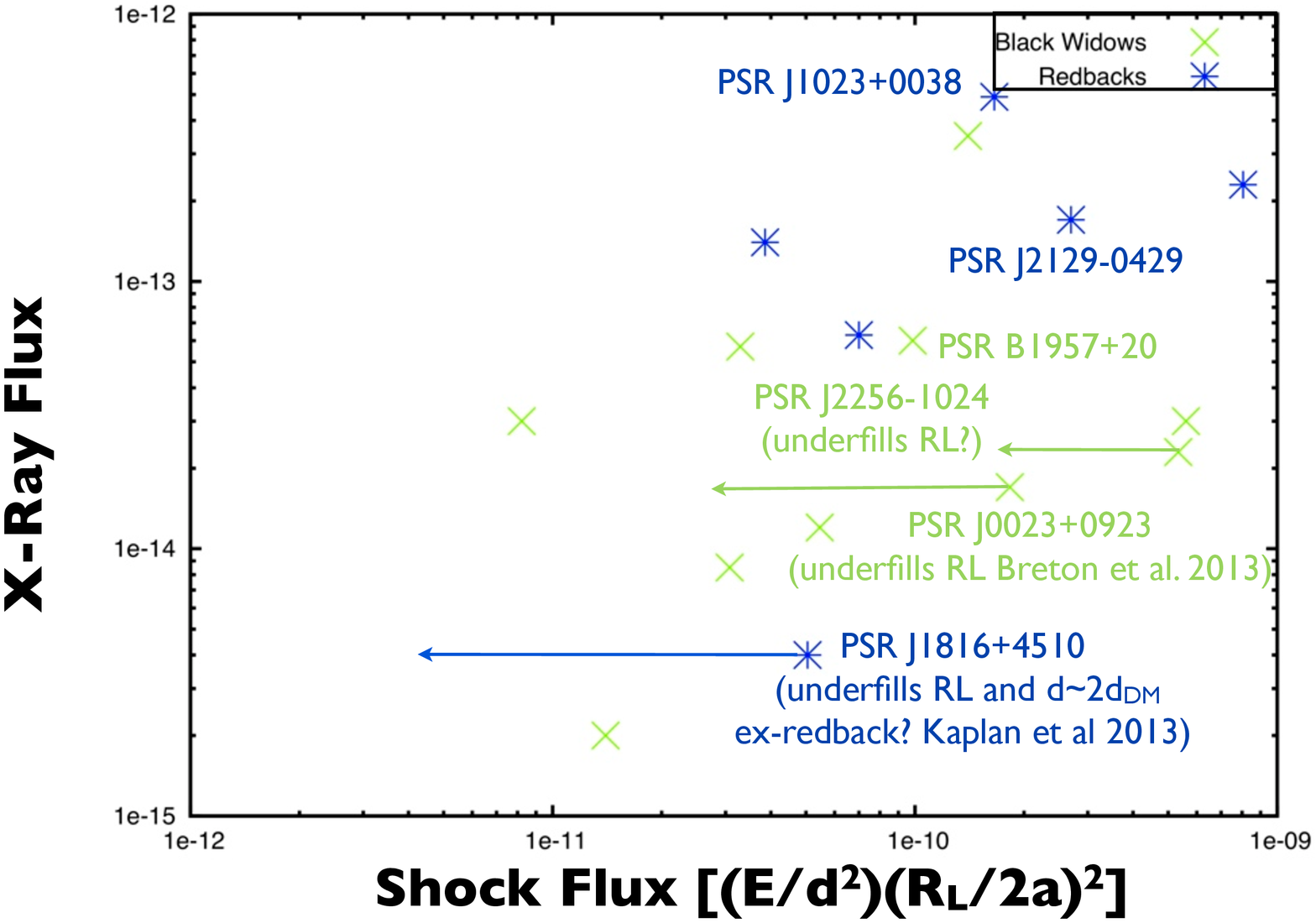}
\caption{0.3-8 keV X-ray flux vs. ``Shock Flux" (see text) for black widows and spiders with X-ray observations, assuming the NE2001 distance (except for the case of PSR J1023+0038) and a Roche lobe filling companion. Arrows point to what positions on the plot certain sources would move to in cases where optical observations suggest the companions  underfill their Roche lobes. }
\label{Figure2}
\end{figure}

Interesting information about the intrabinary shocks in black widows and redbacks as a class is beginning to emerge. Deeper X-ray observations covering multiple orbits and better optical constraints on the companions and the system geometries are needed to model the X-ray light curves and extract information about the pulsar wind. More TeV constraints and a clear detection of orbital modulation at GeV energies would also be very helpful, as would detections at radio wavelengths. These systems give us a different view into the ultra-relativistic winds accelerated in pulsars and the shocks they produce that complements studies of pulsar wind nebulae around young, isolated pulsars.

%%%
%%% -MWL- 2006-01-13 auf Verlagswunsch wieder altes Bibliographie-Format
%%% 
%%%%\subsection{References}
\subsection{Acknowledgements}
Portions of this work were supported by NASA's Fermi Guest Investigator Program, including grant number NNG10PB13P and the Chandra Guest Observer Program including grant numbers GO1-12061B and GO2-13056X.

\end{document}